\documentclass[12pt,preprintnumbers]{article}
\usepackage[utf8]{inputenc}
\usepackage{amsmath, amssymb, wasysym, slashed, multirow,hyperref}
\usepackage{pdfpages}
\usepackage{cite}
\usepackage{dsfont}
\usepackage{times}
\usepackage[export]{adjustbox}

\usepackage{slashed}

\newcommand{\bitem}{\begin{itemize}}
\newcommand{\eitem}{\end{itemize}}
\newcommand{\bwt}{\begin{widetext}}
\newcommand{\ewt}{\end{widetext}}
\newcommand{\beq}{\begin{equation}}
\newcommand{\eeq}{\end{equation}}

\newcommand{\bdm}{\begin{displaymath}}
\newcommand{\edm}{\end{displaymath}}
\newcommand{\bea}{\begin{eqnarray}}
\newcommand{\eea}{\end{eqnarray}}

\renewcommand{\a}{\alpha}

\newenvironment{sciabstract}{%
\begin{quote} }
{\end{quote}}

\newcounter{lastnote}

\usepackage{fancyhdr}
\fancypagestyle{plain}{%
	\fancyhead[R]{RBI-ThPhys-2020-46}

}

\title{Quark mixing with soft breaking of the parity in the minimal Left-Right model}

\author
{Alessio Maiezza \\
\\
\normalsize{Ruder Bo\v skovi\'c Institute, Bijeni\v cka cesta 54, 10000, Zagreb, Croatia}\\ \\
\\
\small{ E-mail: amaiezza@irb.hr}
}

\date{}

\begin{document}


\baselineskip16pt %24 is the original


\maketitle


\begin{sciabstract}
We study the possibility of soft breaking effects of the generalized parity within the minimal Left-Right model.
One aim of the paper is to elaborate on the potentiality, the limit, and the predictivity of a restored parity at high scale.
While revisiting the issue of strong CP in the Left-Right theories, we motivate the possibility of
explicit-parity-breaking, that we then parameterize in the right-handed quark mixing matrix. The strong CP parameter $\bar{\theta}$ is also
parameterized in terms of the breaking. We discuss some possible phenomenological consequences in this
scenario. In particular, the constraint provided by $\bar{\theta}$ enables us to quantify
the maximal deviation of the right-handed quark mixings from the standard case with exact parity in term of a single parameter.
This deviation has a direct impact on flavor physics.
\end{sciabstract}

\section{Introduction}

The asymmetric chiral structure of the weak interactions characterized them apart from the other forces described by the standard model (SM). This asymmetry has motivated
a long and wide route of research, from mirror fermions~\cite{PhysRev.104.254} to Left-Right (LR) theories~\cite{Pati:1974yy,Mohapatra:1974hk,Mohapatra:1974gc,Senjanovic:1975rk}. In the latter, instead of duplicating the whole fermion spectrum, one doubles the weak gauge bosons, and one may hope that this better economy leads to a predictive theory. This opportunity has perfectly seized by the minimal Left-Right symmetric model (MLRSM)~\cite{Senjanovic:1975rk,Senjanovic:1978ev}. It has arisen as a predictive theory of neutrino mass connecting Majorana and Dirac masses~\cite{Nemevsek:2012iq,Senjanovic:2016vxw,Senjanovic:2018xtu,Helo:2018rll}. The MLRSM naturally  embeds the see-saw mechanism~\cite{Minkowski:1977sc,Mohapatra:1979ia} and provides a novel contribution to the neutrinoless double-beta decay~\cite{Mohapatra:1980yp,Tello:2010am} (for a recent study see Ref.~\cite{Li:2020flq}).

There is a deep connection between this process and collider phenomenology~\cite{Dev:2013vxa}.
The possibility of a direct detection at LHC, for example via the Keung-Senjanovi\'c process~\cite{Keung:1983uu}, has generated a renewed interest for the MLRSM. This possibility, because of the minimality of the model, is driven by the low energy constraints. It has been known that flavor physics strictly bounds the model~\cite{Beall:1981ze,Ecker:1985vv}. In the LHC era, detailed studies were done on the possibility to have a low LR scale~\cite{Zhang:2007da,Maiezza:2010ic}.
The bound was refined with several steps and combining different observables in Refs.~\cite{Bertolini:2012pu,Bertolini:2013noa,Bertolini:2014sua,Maiezza:2014ala}, establishing
a low LR scale around 3 TeV. It is disfavoured by perturbative issues in the scalar sector~\cite{Guadagnoli:2010sd,Maiezza:2016bzp,Maiezza:2016ybz,Chauhan:2018uuy}, however, a chance for discovery at LHC survives for a scale around 7 TeV~\cite{Nemevsek:2018bbt}. This benchmark is consistent with the bounds in the recent up-to-dated phenomenological analysis~\cite{Bertolini:2019out}.

Particularly important for the present work are the following concepts and references. First, in Ref.~\cite{Mohapatra:1978fy} the restored parity in LR theories was proposed as a solution to the strong CP problem, in alternative to Peccei-Quinn (PQ) mechanism~\cite{Peccei:1977hh,Weinberg:1977ma,Wilczek:1977pj}. Detailed analysis of strong parity invariance in the MLRSM was performed in Ref.~\cite{Maiezza:2014ala}. Second, in the MLRSM the strong CP is fruitfully predicted by the right-handed quark mixing. In this regard, the RH analogous of the CKM matrix was recently calculated by Senjanovi\'c and Tello (ST) in closed form~\cite{Senjanovic:2014pva,Senjanovic:2015yea}~\footnote{A substantial effort was done also in Ref.~\cite{Zhang:2007da} to get an analytic expression for this matrix, but the approximation used does not capture properly the parametric space.}.

In this paper, we go again through the issue of the strong CP and the restored parity at high energy scale. We elaborate on the possibility of small explicit breaking of the parity together with the standard spontaneous one. In particular, we propose a parametrization of parity-breaking-effects into the right-handed quark mixing, which is determinant for the phenomenology. As an aside, we reproduce the leading ST matrix in a straightforward approach and this may serve to elucidate the original result of Ref.~\cite{Senjanovic:2014pva}. In addition, we parameterize the strong CP parameter $\bar{\theta}$ as well and show how, imposing the severe experimental limit on it, one can describe the possible deviation of the RH quark mixing from the ST matrix.

\section{Main ingredients of the MLRSM}

The MLRSM is based on the gauge group $SU(2)_L\times SU(2)_R\times U(1)_{B-L}\times SU(3)_c$ and an
additional discrete symmetry that exchanges $L \leftrightarrow R$. Left-handed (LH) and Right-handed (RH) fermions
belong to the fundamental representations of $SU(2)_{L,R}$, and in particular, we will work with
the quarks $Q_{L,R }= \left( u \ d \right)^t_{L,R}$. The electric charge is
$Q = I_{3 L} + I_{3 R} + {B - L \over 2}$, being $I_{3 L,R}$ the third generators of
$SU(2)$ groups. The gauge group is broken to the SM one at the energy scale $v_R$, being this the VEV developed by the
neutral component of the RH triplet $\Delta_R (1_L,3_R,2)$. The gauge symmetry is finally broken to $U(1)_{em}$ at electro-weak scale $v$ from the bi-doublet $\Phi (2_L,2_R,0)$~\cite{Senjanovic:1978ev}:
\begin{equation}\label{HH}
\Delta_{R} = \left[ \begin{array}{cc} \Delta^+ /\sqrt{2}& \Delta^{++} \\
\Delta^0 & -\Delta^{+}/\sqrt{2} \end{array} \right]_{R}\,\,\,\,\,\,\,\,
\Phi = \left[\begin{array}{cc}\phi_1^0&\phi_2^+\\\phi_1^-&\phi_2^0\end{array}\right]\,,
\end{equation}
with $\langle\Phi\rangle = \text{diag}\left\{v_1, e^{i \a} v_2 \right\}= v \text{diag}\left\{c_\beta, e^{i \a} s_\beta \right\} $,  $v^2=v_1^2+v_2^2$.
Form here on, we denote $(\sin x, \cos x, \tan x)\equiv (s_x, c_x, t_x)$ being $x$ any angle.

The discrete symmetry can be realized in two alternative ways, either a generalized parity $\mathcal{P}$, considered in this work, or
a generalized charge conjugation $\mathcal{C}$. The former is the original one and is defined as
\begin{equation}\label{Psym}
\mathcal{P}: \left\{ \begin{array}{l} Q_L\leftrightarrow Q_R     \\[1ex]  \Phi \to \Phi^\dagger  \end{array}  \right. \,.
\end{equation}
The generalized parity imposes $g_L=g_R=g$, where $g_{L,R}$ are the couplings of $SU(2)_{L,R}$ and $g$ denotes the standard electro-weak coupling.

The hadronic Yukawa lagrangian reads
\begin{equation}\label{Yukawa}
{\cal L}_Y  = \overline{Q}_L  \Big(Y\, \Phi \,  + \tilde Y \, \tilde\Phi\Big) Q_R + h.c. \,
\end{equation}
with $\tilde\Phi\equiv \sigma_2 \Phi \sigma_2$ being $\sigma_2$ the second Pauli's matrix. When the VEVs are
replaced in Eq.~\eqref{Yukawa}, up and down quark mass terms are generated:
\begin{eqnarray}
M_u&=& v ( c_\beta\, Y  + s_\beta\,{\rm e}^{-i\a}\, \tilde Y ) \nonumber\\
M_d&=& v ( s_\beta \,{\rm e}^{i\a}\, Y + c_\beta\, \tilde Y )\,. \label{masses}
\end{eqnarray}
Writing the quarks in the physical base $Q_{L/R}\rightarrow Q'_{L/R}$, one bi-diagonalizes $M_{u,d}$
\begin{align}
& m_u= U_L^\dag M_u U_R     \nonumber\\
& m_d= D_L^\dag M_d D_R
\end{align}
where $m_{u,d}=m_{u,d \, i}$, $i=1,3$ are the physical quark up and down masses.

In MLRSM the LH and RH currents are the mirror of each other, but the latter are mediated by a heavier twin $W_R$ (with mass $M_{W_R}=g v_R$) of the standard $W_L$:
\begin{align}
\mathcal{L}_{cc}= \frac{g}{\sqrt{2}} W_{L/R}^\mu \, \bar{u}_{L/R} \, \gamma_\mu \, d_{L/R} & =  \frac{g}{\sqrt{2}} W_{L/R}^\mu \, \bar{u}'_{L/R} \, \gamma_\mu \, \underbrace{U_{L/R}^\dag D_{L/R}} \, d'_{L/R}=  \nonumber    \\
& = \frac{g}{\sqrt{2}} W_{L/R}^\mu \, \bar{u}'_{L/R} \, \gamma_\mu \, \underbrace{V_{L/R}} \, d'_{L/R}\,,     \label{CC}
\end{align}
having defined as usual
\begin{equation}\label{CKM}
V_{L/R}=U_{L/R}^\dag D_{L/R} \,,
\end{equation}
where $V_L$ is the standard CKM matrix and $V_R$ is the RH analogous, which we will denote as ST matrix (when $\mathcal{P}$ is only spontaneously broken).

For later convenience, we define the unitary matrices
\begin{equation}\label{definition}
S_u= U_L^\dag U_R \,\,\,\,\,\,\,\,\,\,\,\,\, S_d= D_L^\dag D_R\,\,\,\,\,\,\,\,\,\,\,\,\, S=V_L^\dag V_R        \,.
\end{equation}
From Eqs.~\eqref{CKM} and~\eqref{definition}, $V_R$ can be written as
\begin{equation}\label{e3}
V_R=S_u^\dag V_L S_d    \,.
\end{equation}

\section{Intermezzo: to $\mathcal{P}$ or not to $\mathcal{P}$}

In this section, we argue on the possibility of explicit $\mathcal{P}-$breaking and then motivate on the parametrization of such breaking in the RH quark mixing $V_R$.

\paragraph{Strong CP.} We wish to argue here on possible $\mathcal{P}-$breaking
related to strong CP. We should mention again that a restored $\mathcal{P}$ at high scale was proposed as a solution
of the so-called strong CP problem~\cite{Barr:1991qx,Kuchimanchi:2018ebf,Mimura:2019yfi}. In Ref.~\cite{Maiezza:2014ala}, explicit
$\mathcal{P}-$breaking was also discussed as a possible way-out from the stringent constraint implied on the LR scale.

We go again through the issue of strong CP but in a different light, namely discussing whether one has the right to set to zero
$\theta_{QCD}$ using a restored $\mathcal{P}$. We argue that the argument may not be robust.
Let us think of the quantum mechanically example of instantons in the double-well potential (see for example
Ref.~\cite{Coleman:1978ae}). It is by
construction even, say $Z_2$-symmetric, nonetheless, the instantonic-tunneling-effects break the vacua degeneracy by a factor
$exp(-S/\hbar)$ (being $S$ the classical Euclidean action). Therefore, at the quantum level, the initial symmetry is somehow spoiled.
A similar thing happens with periodic potentials, in the place of the double-well. Notice now that the textbook
narrative teaches us that the vacuum of the Yang-Mills model resembles indeed the one of periodic potential in quantum mechanics.
Our message is that a global $\mathcal{P}$ (in analogy with the $Z_2$ example above) cannot predict a perfectly symmetric quantum vacuum,
although this choice is fully consistent with the restored parity.

\vspace{0.4cm}

We stress the point in the rest of this paragraph, albeit the discussion relies on well-known concepts,
because we think that there is still a lack of clearness in the literature on the Left-Right models.

Consider, as usual, a $SU(2)$
Yang-Mills model with gauge field $A_\mu$ and coupling $g$. The (Euclidean) action is minimized at $8 \pi^2/g^2 n$, being $n$ the winding number (see for example
Ref.~\cite{cheng1984gauge}):
\begin{equation}\label{winding}
n= \frac{1}{16 \pi^2} \int d^4x F_{\mu\nu} \tilde F^{\mu\nu} \,.
\end{equation}
The instanton connects two vacuua characterized by two winding numbers ($p,q$) that differ by unit:
\begin{equation}\label{p_q}
\langle p | e^{-i H t } | q \rangle = \int DA_{n=p-q} \exp \left[ -i (\mathcal{L}+ J A) d^4x \right]\,.
\end{equation}
The true vacuum $\theta$ is a superposition of all with different winding numbers $| \theta \rangle= \sum_n e^{-i n \theta} | n \rangle $ and thus:
\begin{equation}
\langle \theta | e^{-i H t } | \theta \rangle = \sum_n e^{-i n \theta} \int DA_{n} \exp \left[ -i (\mathcal{L}+ J A) d^4x \right] = \sum_n  \int DA_{n} \exp \left[ -i (\mathcal{L}_{eff}+ J A) d^4x \right]\,
\end{equation}
having used Eqs.~\eqref{winding},~\eqref{p_q} and defined $\mathcal{L}_{eff}= \mathcal{L}+ \theta/ (16\pi^2) F_{\mu\nu} \tilde F^{\mu\nu}$.

The \emph{bottom line} is that the topological term reflects an emerging effect. One may still assume that
$\mathcal{P}$ also operates on the quantum vacuum determining the parity-symmetric one, i.e. $\theta=0$.
However, in this case, one pretends that the restored symmetry works well beyond the classical lagrangian and with absolute precision - one then proclaims $\mathcal{P}$ as fundamental as a gauge symmetry. It means that $\mathcal{P}$ has to hold at any arbitrary large energy scale, and such a requirement does not suit well with the absence of a UV completion (see also next paragraph). We interpret the above arguments as a suggestion that the QCD $\theta-$ should be not considered \textit{a priori} vanishing: $\theta=0$ is not a prediction of $\mathcal{P}$ even though the choice is perfectly consistent with it.

Finally, it is worth to recall that, in presence of quark masses, the $\theta-$term can be parameterized via chiral transformations in the non-hermiticity of these masses~\cite{PhysRevLett.42.1195}. For the MLRSM, this implies non-hermitian corrections to the Yukawa couplings. Thus the physical parameter $\bar{\theta}$ can be written as
\begin{equation}
\bar{\theta} = \text{Arg Det} M_u M_d  \,,
\end{equation}
where $M_{u,d}$ in Eq.~\eqref{Yukawa} depend by non-hermitian $Y,\tilde Y$.

\paragraph{UV completion.}
Stating that a discrete symmetry is exact is equivalent to say that such symmetry is gauged. It means that the discrete symmetry is
a remanent part of continuous gauge symmetry, spontaneously broken. It might be the case for the generalized charge conjugation $\mathcal{C}$
if this is UV-completed by $SO(10)$ gauge symmetry (see the discussion in Ref.~\cite{Maiezza:2010ic}). Conversely, there is no way to organize multiplet containing fermions with opposite chirality and, to date, a completion for $\mathcal{P}$ is unknown. It is thus conceivable to expect explicit breaking $\mathcal{P}$, suppressed by some higher energy scale. It might be theoretically interesting to speculate whether, by orbifolding in higher space-time dimensions (with some technique along the line of Ref.~\cite{Sundrum:2005jf}), one could complete $\mathcal{P}$ in an analogous
manner in which $\mathcal{C}$ is completed in 4D. However, going back to 4D one would deal with $\mathcal{P}-$breaking operator suppressed as powers of $v_R/v_{comp}$, denoting with $v_{comp}$ some compactification scale.

Pragmatically, one can invoke effective operators that break the generalized parity.
Within MLRSM, the leading one is dimension six~\cite{Maiezza:2014ala}:
\begin{equation}
\frac{1}{M} \overline{Q}_L\, Y\, \Phi \, Q_R \,Tr \left(\Delta_R^\dagger \Delta_R \right)\,,
\end{equation}
with $Y$ now non-hermitian and $M$ being any high energy scale. The correction to the non-hermiticity of the Yukawa couplings is $v_R^2/M^2$. If one identifies
$M$ conservatively with Plank scale and $v_R$ in the TeV range, as in Ref.~\cite{Maiezza:2014ala}, the correction is negligible (order $10^{-30}$). However, assuming that the Plank scale is the cut-off for $\mathcal{P}$, although a possibility, may not be well-motivated; any eventual field beyond MLRSM might be a singlet of $\mathcal{P}$.

\section{Righ-handed mixing matrix}\label{leading}

Motivated by the discussions above, in this central section we parameterize soft breaking of $\mathcal{P}$ symmetry into the RH quark mixing matrix $V_R$.

Let us start by inverting Eqs.~\eqref{masses}
\begin{align}
& Y= \frac{c_\beta}{v c_{2\beta}} \left( M_u-t_\beta e^{-i \alpha} M_d \right)  \nonumber   \\
& \tilde Y=\frac{c_\beta}{v c_{2\beta}} \left( M_d-t_\beta e^{i \alpha} M_u \right)    \label{YY}    \,,
\end{align}
and consider the equations
\begin{align}
& Y-Y^\dag=i k A  \nonumber   \\
& \tilde Y-\tilde Y^\dag=i k B  \label{nonH}   \,,
\end{align}
where $A,B$ are hermitian matrices and $k$ is a formally small expansion parameter. If $k=0$, one recovers the case of exact $\mathcal{P}$-symmetry, otherwise one parameterizes an explicit breaking of this generalized parity. Moreover, one can parameterizes the suppression only in $k$ and the largest matrix elements of $A$ or $B$ is unit in absolute value.

We define the dimensional parameter
\begin{equation}
\bar{v} =  \frac{v c_{2\beta}}{c_{\beta} }   \,.
\end{equation}
Consider now the first of Eqs.~\eqref{nonH}
and replace the expression of $Y$ from Eqs.~\eqref{YY}. By multiplying on the left for $U_L^\dag$ and on the right for $U_R$, one obtains
\begin{equation}
m_u-S_u m_u S_u -t_\beta e^{-i \alpha}\, V_L m_d V_R^\dag + t_\beta e^{i \alpha}\, S_u V_R m_d V_L^\dag S_u= i k \bar{v}\, U_L^\dag A U_R  \,,
\end{equation}
where the relations in Eqs.~\eqref{CKM} and~\eqref{definition} are used. Employing again these relations, the R.H.S. of this expression can be written as
\begin{equation}\label{smallA}
i k \bar{v} \, \underbrace{U_L^\dag A U_L} S_u= i k \bar{v} \, \underbrace{a} S_u   \,,
\end{equation}
where $a$ is diagonal, and in the last step it has been used the freedom to diagonalize the hermitian matrix $A$ in terms of the unitary transformation $U_L$.
In this way, one arrives to the expression
\begin{equation}\label{e1}
m_u-S_u m_u S_u -t_\beta e^{-i \alpha} \,  V_L m_d V_R^\dag + t_\beta e^{i \alpha} \,  S_u V_R m_d V_L^\dag S_u= i k \bar{v} \,  a S_u   \,.
\end{equation}
Doing the equivalent manipulation on the second of Eqs.~\eqref{nonH}(but multiplying now on the left for $D_L^\dag$ and on the right for $D_R$), one gets
\begin{equation}\label{e2}
m_d-S_d m_d S_d -t_\beta e^{i \alpha} \, V_L^\dag m_u V_R + t_\beta e^{-i \alpha} \,  S_d V_R^\dag m_u V_L S_d= i k \bar{v} \, B S_d   \,.
\end{equation}
where now $B$ cannot be chosen diagonal (because $D_L=U_L V_L$, $V_L$ is fixed and $U_L$ has been already chosen), and we have recalled $(D_L^\dag B D_L) \rightarrow B$ since $B$ is an arbitrary hermitian matrix.

Replacing Eq.~\eqref{e3} into Eqs.~\eqref{e1} and~\eqref{e2}, one gets
\begin{align}\label{system}
&  m_u -S_u m_u S_u -t_\beta e^{-i \alpha} \, V_L m_d S_d^\dag V_L S_u +t_\beta e^{i \alpha} \, V_L S_d m_d V_L^\dag S_u   =     i k \bar{v} \, a S_u   \nonumber      \\
&  m_d -S_d m_d S_d -t_\beta e^{i \alpha} \, V_L^\dag m_u S_u^\dag V_L S_d +t_\beta e^{-i \alpha} \, V_L^\dag S_u m_u V_L S_d   =     i k \bar{v} \, B S_d   \,.
\end{align}
This is a system to be solved for the unknown matrices $S_{u,d}$ and then, once the solution is at hand, one reconstructs $V_R$ via Eqs.~\eqref{CKM} and~\eqref{definition}.

\paragraph{The double expansion}

We want to solve perturbatively the Eqs.~\eqref{system}. To this aim, note that if $y\equiv t_\beta sin \alpha=0= k$ all the matrices in Eq.~\eqref{definition} reduce to the identity and thus $V_R=V_L$. Therefore, it is convenient to parameterize all the matrices as a power series in both $y,k$.

Let us write a generic hermitian matrix $\mathcal{H}$ as:
\begin{equation}\label{Hexp}
\mathcal{H}=\sum_{r,p} k^r y^p H_{r,p}\,.
\end{equation}
The needed double-expansion for an unitary matrix $U$ can be obtained by replacing Eq.~\eqref{Hexp} in $U=e^{i \mathcal{H}}= \sum_{n=0}^{\infty} (i\mathcal{H})^n/n!$ so,
truncating for example at second order, one gets
\begin{equation}
U=1+i \left(y H_{0,1}+k  H_{1,0}\right)+i \left(y^2 H_{0,2}+y k  H_{1,1}+k^2 H_{2,0}\right)-\frac{1}{2} \left(y^2 H_{0,1}^2+2 y k  H_{1,0} H_{0,1}+k^2 H_{1,0}^2\right)\,.
\end{equation}
The matrix $U$ indicates either $S_u,S_d,S$ with $H=H_u,H_d,H_R$ respectively. Therefore, also the matrix $V_R= V_L S$ is parameterized in terms of a power series in $k$ and $y$.

Plugging the expansions in Eqs.~\eqref{system} one gets linear systems order by order in powers of $k,y$.

\paragraph{Leading contribution to $V_R$ due to the explicit $\mathcal{P}$-breaking.}
We start solving the linear system obtained by taking the coefficients $k^1, y^0$ in the Eqs.~\eqref{system}. One gets:
\begin{align}
&  m_u H_{u\, 1,0}+H_{u\, 1,0} m_u + \bar{v} a = t_{\beta } \left(V_L m_d H_{d\, 1,0} V_L^{\dagger }+V_L H_{d\, 1,0} m_d V_L^{\dagger }\right) \nonumber \\
&  m_d H_{d\, 1,0}+H_{d\, 1,0} m_d + \bar{v} B = t_{\beta } \left( V_L^{\dagger } m_u H_{u\, 1,0} V_L + V_L^{\dagger} H_{u\, 1,0}  m_u V_L \right) \label{linear_system_P_broken} \,.
\end{align}
Since $m_{u,d}$ are diagonal, the solution for $H_{u\, 1,0},H_{d\, 1,0}$ is simply:
\begin{align}
& \left[H_{u\, 1,0}\right]_{i j} =  -\frac{\bar{v}}{1-t_\beta^2} \frac{\left[a+t_\beta V_L B V_L^\dagger   \right]_{i,j}}{m_{u\, i }+m_{u\, j}}      \\
& \left[H_{d\, 1,0}\right]_{i j} =  \frac{\bar{v}}{1-t_\beta^2} \frac{\left[B-t_\beta V_L^\dagger a V_L   \right]_{i,j}}{m_{d\, i }+m_{d\, j}} \,,
\end{align}
where the indices $i,j$ indicate explicitly the matrix elements. Finally, by considering the expansion at the same order of Eq.~\eqref{e3} and of the expression $V_R=V_L S$:
\begin{align}
& V_L^\dagger H_{u\, 1,0}+H_{R\, 1,0} V_L^\dagger = H_{d\, 1,0} V_L^\dagger   \nonumber   \\
& V_R= V_L + i k H_{R\, 1,0} \,
\end{align}
one arrives at the desired expression at order $\mathcal{O}(k^2)$
\begin{align}
\left[ V_R^{\slashed P} \right]_{i j} & = \left[ V_L \right]_{ij}+ \frac{i k \bar{v}}{1-t_\beta^2}\left\{ \frac{\left[V_L\right]_{i n} \left[B-t_\beta V_L^\dagger a V_L\right]_{n j}}{m_{d\, n}+m_{d\, j}} +\frac{ \left[a+t_\beta V_L B V_L^\dagger\right]_{i n}  \left[V_L\right]_{n j}}{m_{u\, i}+m_{u\, n}}\right\}   \nonumber   \\
& \equiv V_L + V_R^{cor} \,,   \label{VR_P_broken}
\end{align}
where summation on the index $n$ is understood. Recall also that the matrices $a,B$ are defined in Eqs.~\eqref{nonH} and~\eqref{smallA}. We put on $V_R$ the index $\slashed P$ to remember that this comes only from an explicit $\mathcal{P}$-breaking, and the corresponding correction to CKM matrix has been denoted as $V_R^{cor}$.

\paragraph{Leading contribution to $V_R$ due to the spontaneous $\mathcal{P}$-breaking.}
Taking now the coefficients $k^0, y^1$ in Eqs.~\eqref{system}, the calculation is \textit{mutatis mutandis} as in the previous paragraph, and it has to recover the leading ST matrix in Ref.~\cite{Senjanovic:2014pva}. It is worth noting that the present derivation is complementary to the original one, which relies on the notion of the square-root matrix.

The linear system to be solved is
\begin{align}
&  m_u H_{u\, 0,1}+H_{u\, 0,1} m_u - 2 V_L m_d V_L^\dagger = t_{\beta } \left(V_L m_d H_{d\, 0,1} V_L^{\dagger }+V_L H_{d\, 0,1} m_d V_L^{\dagger }\right) \nonumber \\
&  m_d H_{d\, 0,1}+H_{d\, 0,1} m_d + 2 V_L^\dagger m_u V_L = t_{\beta } \left( V_L^{\dagger } m_u H_{u\, 0,1} V_L + V_L^{\dagger} H_{u\, 0,1}  m_u V_L \right) \label{linear_system_SSB}      \,.
\end{align}
Note that this is equivalent to Eq.~\eqref{linear_system_P_broken} provided the replacements $a \rightarrow -2 V_L m_d V_L^\dagger$ and $B \rightarrow 2 V_L^\dagger m_u V_L$, therefore the solution for $V_R$ at order $\mathcal{O}(y^2)$ can be gotten simply substituting them in Eq.~\eqref{VR_P_broken}:
\begin{equation}
\left[ V_R \right]_{i j} = \left[ V_L \right]_{ij}+ \frac{2 i y}{1-t_\beta^2}\left\{ \frac{\left[V_L\right]_{i n} \left[V_L^\dagger m_u V_L + t_\beta m_d \right]_{n j}}{m_{d\, n}+m_{d\, j}} +\frac{ \left[t_\beta m_u -V_L m_d V_L^\dagger \right]_{i n}  \left[V_L\right]_{n j}}{m_{u\, i}+m_{u\, n}}\right\}   \,.
\end{equation}
Finally, recalling that $y= t_\beta \sin\alpha$ and using that $t_\beta/(1-t_\beta^2)=1/2 t_{2\beta}$, one gets
\begin{equation}\label{ST}
\left[ V_R^{ST} \right]_{i j} = \left[ V_L \right]_{ij}+ i s_\alpha t_{2\beta} \left\{ t_\beta \left[ V_L \right]_{ij} + \frac{\left[V_L\right]_{i n} \left[V_L^\dagger m_u V_L \right]_{n j}}{m_{d\, n}+m_{d\, j}} -  \frac{ \left[ V_L m_d V_L^\dagger \right]_{i n}  \left[V_L\right]_{n j}}{m_{u\, i}+m_{u\, n}}\right\}    \,,
\end{equation}
being the summation over $n$ understood. We put an index $ST$ to remember that this is only the standard contribution due to the spontaneous $\mathcal{P}$-breaking, and indeed it has to be compared with the ST matrix~\cite{Senjanovic:2014pva}. Note that there is an apparent discrepancy in the overall sign in front of $i s_\alpha t_{2\beta}$ and the relative sign inside the curly brackets, however, this is easily traced back to a different definition of the Yukawa lagrangian in Eq.~\eqref{Yukawa} (with respect of Eq.~(1) of Ref.~\cite{Senjanovic:2014pva}) and the VEVs in Eq.~\eqref{HH}.

At leading order, $V_R$ matrix is thus the sum of the contributions in Eqs.~\eqref{ST} and~\eqref{VR_P_broken}:
\begin{equation}\label{VRTOT}
V_R=V_R^{ST}+V_R^{cor}\,.
\end{equation}
Note that the dominant correction to CKM matrix in Eq.~\eqref{ST} is $\approx s_\alpha t_{2\beta} \frac{m_t}{2m_b}$~\cite{Senjanovic:2014pva}, and the dominant correction in Eq.~\eqref{VR_P_broken} is $\approx \frac{\bar{v}}{1-t_\beta^2} \frac{1}{m_u}$. To have a convergent expansion in Eq.~\eqref{VRTOT}, these expressions have to be small:
\begin{equation}\label{radi}
s_\alpha t_{2\beta}<2 \frac{m_b}{m_t}\,,\,\,\,\,\,\,\,\,\,\,\,\, k<\frac{m_u (1-t_\beta^2)}{\bar{v}}\,,
\end{equation}
being the former indeed the condition in Ref.~\cite{Senjanovic:2014pva}.

It is worth recalling that, exactly as discussed in Refs.~\cite{Senjanovic:2014pva,Senjanovic:2015yea}, there is a discrete family of solutions for $V_R$ by replacing in Eqs.~\eqref{ST} and~\eqref{VR_P_broken} (and then in Eq.~\eqref{VRTOT}) $m_{u\,i}\rightarrow s_{u\,i} m_{u\,i}$ and $m_{d\,i}\rightarrow s_{d\,i} m_{d\,i}$, being $s_{u,d}$ arbitrary signs. At higher order, the parametrization of $V_R$ becomes more involved (see Appendix~\ref{higher}), however, the leading expression~\eqref{VRTOT} is enough in the assumption that the explicit breaking is only a corrections to the case with exact $\mathcal{P}$.

\begin{figure}[t]\label{fig_scatter}
\centerline{\includegraphics[width=0.49\textwidth,valign=t]{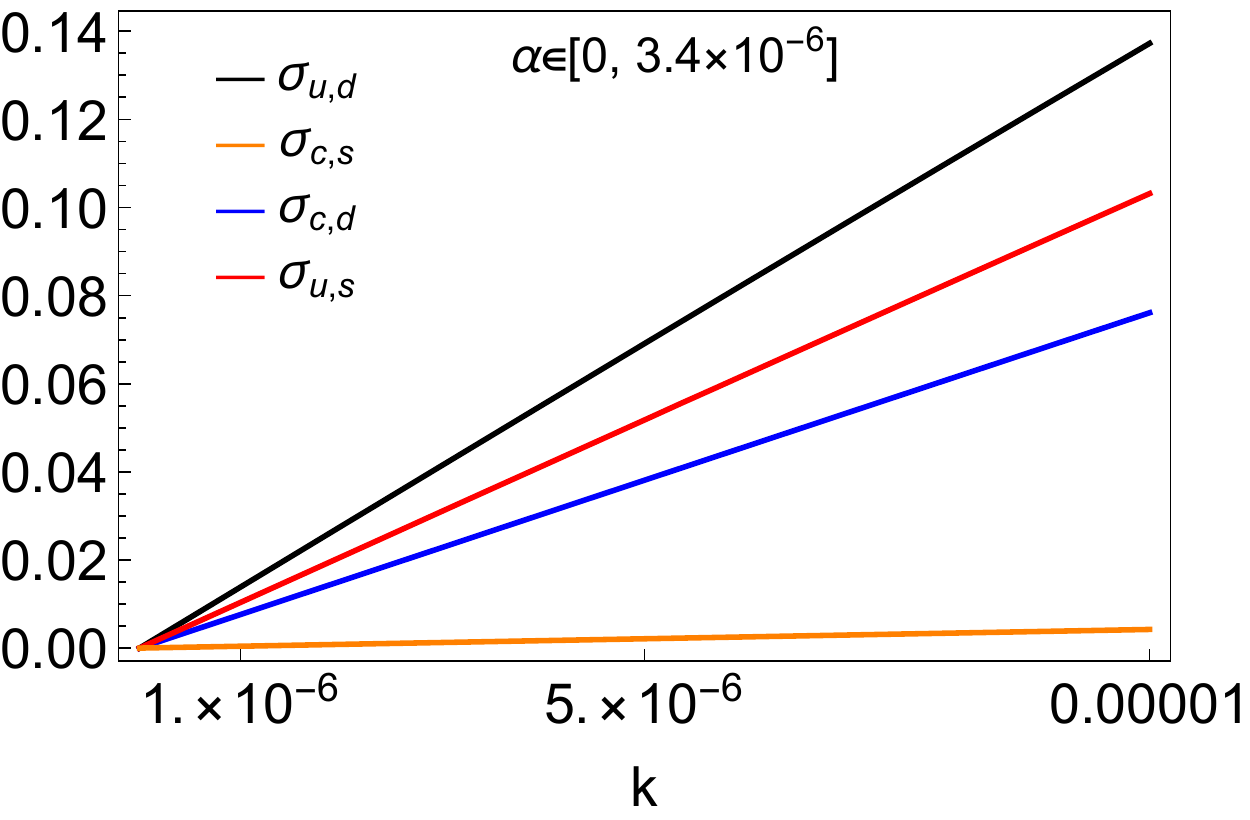}~\includegraphics[width=0.49\textwidth,valign=t]{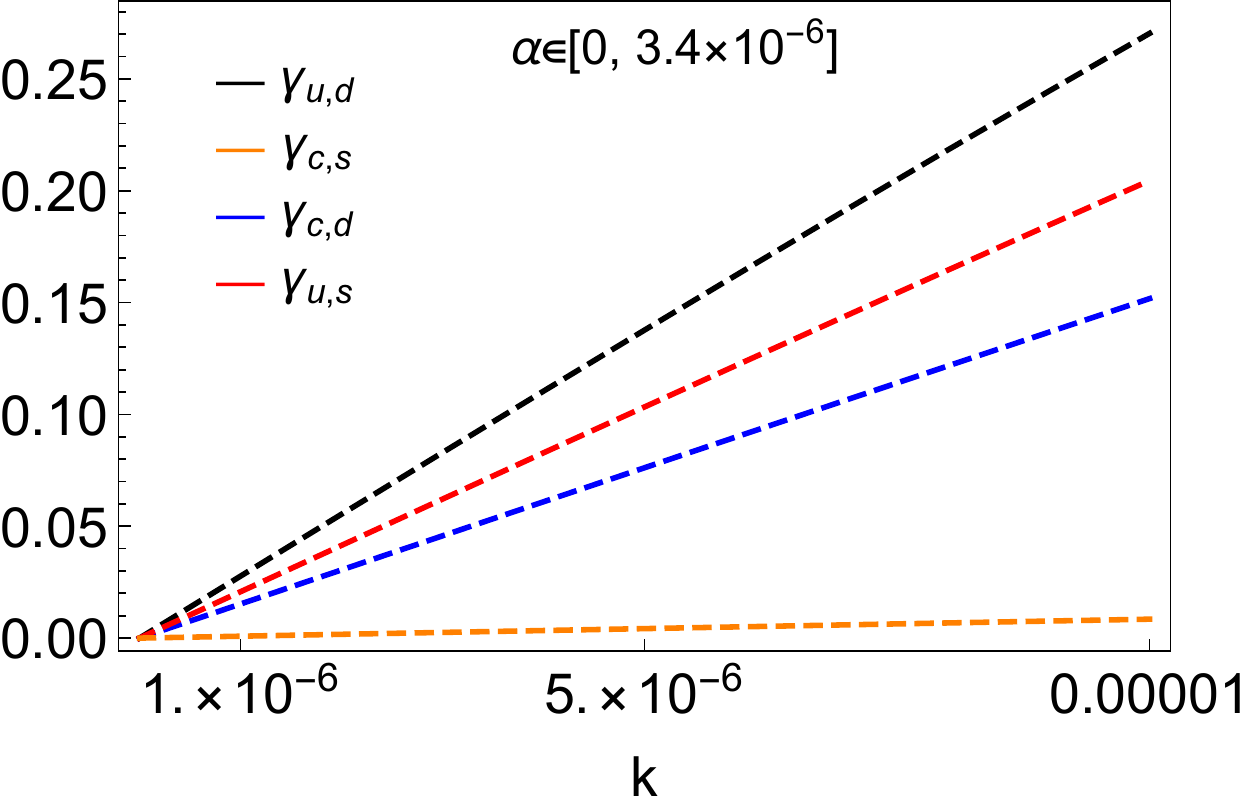}}
\caption{Left panel: $\sigma_{i,j}$ defined in Eq.~\eqref{sigma_gamma} (only Cabibbo sub-matrix shown as an example), while saturating the experimental limit on $\bar{\theta}$. We fix $t_\beta=0.1$ and the range of $k$ is consistent with the limit in Eq.~\eqref{radi}. Right panel: corresponding values for $\gamma_{i,j}$.}
\end{figure}

\section{Discussion and possible phenomenological impact}

The most sensitive observable to both spontaneous and soft $\mathcal{P}-$breaking is the neutron electric dipole moment that provides the severe constraint
\begin{equation}\label{t_limit}
\bar{\theta} = \text{Arg Det} M_u M_d= \text{Arg Det} S_u^2 V_R \lesssim 10^{-10}  \,,
\end{equation}
where we used the expression of $M_{u,d}$ in Eq.~\eqref{masses} (see also Ref.~\cite{Senjanovic:2015yea}). Thus $\bar{\theta}$ depends on $\sin\alpha t_{2\beta}, k$, and the non-hermiticity of the Yukawa couplings. The experimental limit comes from the nEDM~\cite{Ottnad:2009jw}.
Using the expansion of the previous section, the leading expression for $\bar{\theta}$ can be written as
\begin{align}
\bar{\theta} = & \frac{1}{2}\sin \alpha t_{2\beta} Re\, Tr \left[ m_d^{-1} V_L^\dagger m_u V_L + m_u^{-1} V_L m_d V_L^\dagger \right]+ \nonumber   \\
& \frac{k \bar{v}}{2(1-t_\beta^2)} Re\, Tr \left[ m_d^{-1} B -  m_u^{-1} a -t_\beta ( m_d^{-1} V_L^\dagger a V_L +  m_u^{-1}V_L B V_L^\dagger) \right]  \label{tbar}  \,.
\end{align}
The first line is due to the spontaneous $\mathcal{P}-$breaking and reproduces the expression given in Ref.~\cite{Senjanovic:2015yea}, provided the different definition of the Yukawa lagrangian is used, as discussed above; the second line is the correction due to the explicit $\mathcal{P}-$breaking.

A limiting scenario is when there is no cancellation at all between the two contributions in~\eqref{tbar}. In this case, $\sin \alpha t_{2\beta}$ and $k$ are separately constrained to be $\lesssim 10^{-12}$~\footnote{Note that one saturates $\bar{\theta}=10^{-10}$ from values order $\approx 10^{-12}$ for $\sin \alpha t_{2\beta}, k$, gaining two order of magnitude because an enhancement due to $m_t$}. This would mean that $\mathcal{P}$ is just very softly broken and the impact of the breaking effects would be marginal on flavor physics, in particular on $\epsilon,\epsilon'$. The matrix $V_R$ would reduce effectively to $V_L$ up to corrections $\approx  10^{-8}-10^{-12}$ depending on the specific matrix elements. Clearly, this remains true also in the standard case with no explicit breaking of $\mathcal{P}$, and this has led to the severe bound $M_{W_{R}}>20$TeV in Ref.~\cite{Maiezza:2014ala}.

\vspace{0.5cm}

In order to satisfy $\bar{\theta}\lesssim 10^{-10}$, there is the phenomenological possibility of a cancellation between the two lines, i.e. between the term proportional to $\sin \alpha t_{2\beta}$ and the one proportional to $k$. This fine-tuning with a precision of $\approx 10^{-10}$ may be regarded as the "strong CP problem" within the MLRSM with a soft breaking of the generalized parity. There is no \textit{a priori} theoretical reason for this cancellation, but it still may happen. The picture can be summarized graphically as in Fig.~\ref{fig_scatter}. In particular, we show the deviation of $V_R$ from ST matrix as a function of $k$, consistent with the limit in Eq.~\eqref{tbar}. We define
\begin{align}
& \sigma_{i,j} = \left|\frac{\left[V_R\right]_{i\,j}-\left[V_R^{ST}\right]_{i\,j}}{\left[V_R\right]_{i\,j}+\left[V_R^{ST}\right]_{i\,j}}\right|  \nonumber   \\
& \gamma_{i,j} = \left| Arg \frac{\left[V_R\right]_{i\,j}}{\left[V_R^{ST}\right]_{i\,j}} \right|\,,     \label{sigma_gamma}
\end{align}
and maximize the deviation over the arbitrary matrices $a,B$ consistently with the experimental constraint in Eq.~\eqref{t_limit}.
The maximal corrections are delimited by straight lines because of the approximated linearity in both $k$ and $y$ (or $\alpha$) in Eq.~\eqref{VRTOT} -  non-linear corrections are quite suppressed in the interesting regime for which ST matrix is spoiled only up to a few percent.
Both $\sin \alpha t_{2\beta}$ and $k$ can be fairly larger than $\lesssim 10^{-12}$ (which is when no fine-tuning between $k$ and $\alpha$ occurs), and thus the low energy predictions of the MLRSM may be affected by the corrections in Eq.~\eqref{VRTOT}. There is a number of studies in literature analyzing correlated constraints from nEDM together with other CP-violating observables, as $\epsilon,\epsilon'$ (among others, see Refs.~\cite{Xu:2009nt,Haba:2018byj,Maiezza:2014ala,Bertolini:2019out}), but not taking into account the impact of $\mathcal{P}$-breaking on $V_R$.  The parametrization in Eq.~\eqref{VRTOT} could serve in this light.

Because of $\bar{\theta}$ in Eqs.~\eqref{t_limit}and~\eqref{tbar}, small $\mathcal{P}-$breaking does not destroy completely the predictivity of the MLRSM. In this sense, strong CP is still a "blessing", according to the view in Ref.~\cite{Senjanovic:2020int}, for it enables us to constrain the RH quark mixing. However, there is still the possibility to invoke an additional mechanism that cancels the strong CP-term~\cite{Maiezza:2014ala,Bertolini:2019out}. In this case, Eq.~\eqref{tbar} does not longer hold. Clearly, the parametrization of $V_R$ in Eq.~\eqref{VRTOT} is generic and thus still works, but one loses information from strong CP. In this regard, a final comment is in order. The additional mechanism could be realized by any implementation of PQ symmetry within MLRSM. In this case, one should do model-dependent studies, in fact completely abandoning the MLRSM. Nonetheless, also in this hypothetical extended Left-Right scenario, one should still consider the explicit breaking of both $\mathcal{P}$ and PQ symmetry.

\section{Summary}

In this work, we have provided a critical discussion of the generalized parity $\mathcal{P}$ in the MLRSM. We have given a fresh look to strong CP invariance in the light of
a restored parity at high energy. In particular, we have argued that the argument of setting the topological QCD $\theta$ to zero is not robust. The basic concept is that this
anomalous term is not just a coupling among the others in the lagrangian, because it is generated dynamically by the vacuum structure, the instantons, and the barrier penetration. Requiring that the restored parity also selects the symmetric vacuum with absolute precision, is equivalent to promote $\mathcal{P}$ as a gauge symmetry, which is not the state-of-art. Instantons may then induce a $\mathcal{P}-$breaking effect on the quark masses, which manifests itself through a non-hermiticity of the Yukawa couplings.

On a more general ground, we have recalled that $\mathcal{P}$  also may be explicitly broken by high energy physics because the generalized parity seems not to have a UV completion. Usually, one has to assume that the explicit breaking is small in comparison with the spontaneous one.
We have provided a parametrization that enables us to quantify the meaning of this "small". In particular, we have parameterized the RH analogous of the CKM matrix in terms
of the non-hermiticity of the Yukawa couplings that follows from the explicit breaking. As a result, we have written in a closed-form the corrections to the ST matrix, which in turn we have reproduced in a slightly different way, coming from the $\mathcal{P}-$ breaking.

The form of $V_R$ is phenomenologically relevant because it enters into very sensitive CP-violation observables as the EDM of the neutron and $\epsilon'$.
Not less important, $(V_R)_{ud}$ matters for collider physics, being the production of $W_R$ proportional to $|(V_R)_{u,d}|^2$. An experimental detection
at LHC or next colliders would also help to disentangle between scenarios with exact $\mathcal{P}$ or not.

%%%%%%%%%%%%%%%%%%%%%%%%%%%%%%%%%%%%%%%%%%%%%%%%%%%%%%%%%%%%%%%%%%%%%%%%%%%%%%%%%%%%%%%%%%%%%%%%%%%%%%%%%%%%%%%%%%%%%%%%%%%%%%%%%%%%%%%%%%%%%%%%%%%%%%%%%%%%%%%%%%%%%%%%%%%%%%%%%%

\section*{Acknowledgement}

The author thanks Miha Nemev\v{s}ek, Fabrizio Nesti and Goran Senjanovi\'c for discussions and useful comments on the manuscript.
The author was partially supported by the Croatian Science Foundation project number 4418.

%%%%%%%%%%%%%%%%%%%%%%%%%%%%%%%%%%%%%%%%%%%%%%%%%%%%%%%%%%%%%%%%%%%%%%%%%%%%%%%%%%%%%%%%%%%%%%%%%%%%%%%%%%%%%%%%%%%%%%%%%%%%%%%%%%%%%%%%%%%%%%%%%%%%%%%%%%%%%%%%%%%%%%%%%%%%%%%%%%%%

\appendix

\section{Higher order corrections}\label{higher}

The solution for $V_R$ at higher-order proceeds at the same way than the leading one, namely solving linear systems in $H_{u\,n,m},H_{d\,n,m},H_{R\,n,m}$ where $n,m$ set the order being the exponents of the expansion parameters: $k^n\times y^m$. Although one deals at each order just with linear systems, already at second order ($n+m=2$) the expressions become rather cumbersome, as it will be clear below. It is then convenient leaving the parametrization of $V_R$ in an implicit form in terms of equations, as illustrated in what follows. Being linear, these equations are readily evaluated once numerical parameters are put in.

Taking the coefficient of $k\times y$ in the Eqs.~\eqref{system}and~\eqref{e3},  one gets three linear equations to be solved for $H_{u\,1,1},H_{d\,1,1},H_{R\,1,1}$, since the matrices $H_{u\,0,1},H_{d\,0,1},H_{R\,0,1}$ and $H_{u\,1,0},H_{d\,1,0},H_{R\,1,0}$ are known from the leading solution in Sec.~\ref{leading}:
\begin{align}
& \frac{1}{2} m_u H_{u\, 1,0} H_{u\, 0,1}+H_{u\, 0,1} m_u H_{u\, 1,0}+\frac{1}{2} H_{u\, 0,1} H_{u\, 1,0} m_u+H_{u\, 1,0} m_u H_{u\, 0,1} + \nonumber  \\
& \frac{1}{2} H_{u\, 1,0} H_{u\, 0,1} m_u+ V_L m_d H_{d\, 1,0} V_L^\dagger-V_L H_{d\, 1,0} m_d V_L^\dagger-2 V_L m_d V_L^\dagger H_{u\, 1,0} +   \nonumber  \\
& \bar{v} a H_{u\, 0,1}-i m_u H_{u\, 1,1}-i H_{u\, 1,1} m_u+\frac{1}{2} m_u H_{u\, 0,1} H_{u\, 1,0} +  \nonumber  \\
&  t_\beta \left(  -V_L m_d H_{d\, 0,1} V_L^\dagger H_{u\, 1,0}-V_L m_d H_{d\, 1,0} V_L^\dagger H_{u\, 0,1}-V_L H_{d\, 0,1} m_d V_L^\dagger H_{u\, 1,0} -  \right.  \nonumber  \\
& V_L H_{d\, 1,0} m_d V_L^\dagger H_{u\, 0,1}+i V_L m_d H_{d\, 1,1} V_L^\dagger+i V_L H_{d\, 1,1} m_d V_L^\dagger + \nonumber  \\
&  \frac{1}{2} V_L m_d H_{d\, 0,1} H_{d\, 1,0} V_L^\dagger+\frac{1}{2} V_L m_d H_{d\, 1,0} H_{d\, 0,1} V_L^\dagger-\frac{1}{2} V_L H_{d\, 0,1} H_{d\, 1,0} m_d V_L^\dagger-    \nonumber  \\
& \left. \frac{1}{2} V_L H_{d\, 1,0} H_{d\, 0,1} m_d V_L^\dagger \right)=0   \\
& \frac{1}{2} m_d H_{d\, 1,0} H_{d\, 0,1}+H_{d\, 0,1} m_d H_{d\, 1,0}+\frac{1}{2} H_{d\, 0,1} H_{d\, 1,0} m_d+H_{d\, 1,0} m_d H_{d\, 0,1}+ \nonumber   \\
& \frac{1}{2} H_{d\, 1,0} H_{d\, 0,1} m_d +2 V_L^\dagger m_u V_L H_{d\, 1,0}-V_L^\dagger m_u H_{u\, 1,0}.V_L+V_L^\dagger H_{u\, 1,0} m_u V_L +  \nonumber   \\
&  \bar{v} B H_{d\, 0,1}-i m_d H_{d\, 1,1}-i H_{d\, 1,1} m_d+\frac{1}{2} m_d H_{d\, 0,1} H_{d\, 1,0} + \nonumber   \\
& t_\beta \left( -V_L^\dagger m_u H_{u\, 0,1} V_L H_{d\, 1,0}-V_L^\dagger m_u H_{u\, 1,0} V_L H_{d\, 0,1}-V_L^\dagger H_{u\, 0,1} m_u V_L H_{d\, 1,0} - \right. \nonumber   \\
& V_L^\dagger H_{u\, 1,0} m_u V_L H_{d\, 0,1}+i V_L^\dagger m_u H_{u\, 1,1} V_L+i V_L^\dagger H_{u\, 1,1} m_u V_L +  \nonumber   \\
&  \frac{1}{2} V_L^\dagger m_u H_{u\, 0,1} H_{u\, 1,0} V_L+\frac{1}{2} V_L^\dagger m_u H_{u\, 1,0} H_{u\, 0,1} V_L-\frac{1}{2} V_L^\dagger H_{u\, 0,1} H_{u\, 1,0} m_u.V_L-   \nonumber   \\
& \left. \frac{1}{2} V_L^\dagger H_{u\, 1,0} H_{u\, 0,1} m_u V_L \right)=0 \\
& -H_{d\, 0,1} H_{R\, 1,0} V_L^\dagger-H_{d\, 1,0} H_{R\, 0,1} V_L^\dagger-i H_{d\, 1,1} V_L^\dagger+\frac{1}{2} H_{d\, 0,1} H_{d\, 1,0} V_L^\dagger+  \nonumber   \\
& \frac{1}{2} H_{d\, 1,0} H_{d\, 0,1} V_L^\dagger+i V_L^\dagger H_{u\, 1,1}-\frac{1}{2} V_L^\dagger H_{u\, 0,1} H_{u\, 1,0}  -  \nonumber   \\
&  \frac{1}{2} V_L^\dagger H_{u\, 1,0} H_{u\, 0,1}+i H_{R\, 1,1} V_L^\dagger+\frac{1}{2} H_{R\, 0,1} H_{R\, 1,0} V_L^\dagger+\frac{1}{2} H_{R\, 1,0} H_{R\, 0,1} V_L^\dagger=0
\end{align}
Next, taking the coefficient of $k^2\times y^0$ in the Eqs.~\eqref{system}and~\eqref{e3},  one gets three linear equations to be solved for $H_{u\,2,0},H_{d\,2,0},H_{R\,2,0}$:
\begin{align}
& \bar{v} a H_{u\, 1,0}-i m_u H_{u\, 2,0}-i H_{u\, 2,0} m_u+\frac{1}{2} m_u H_{u\, 1,0} H_{u\, 1,0} +  \nonumber   \\
& H_{u\, 1,0} m_u H_{u\, 1,0}+\frac{1}{2} H_{u\, 1,0} H_{u\, 1,0} m_u+ t_\beta \left( -V_L m_d H_{d\, 1,0} V_L^\dagger H_{u\, 1,0}\right. -   \nonumber   \\
& V_L H_{d\, 1,0} m_d V_L^\dagger H_{u\, 1,0}+i V_L m_d H_{d\, 2,0} V_L^\dagger+i V_L H_{d\, 2,0} m_d V_L^\dagger +   \nonumber   \\
& \left. \frac{1}{2} V_L m_d H_{d\, 1,0} H_{d\, 1,0} V_L^\dagger-\frac{1}{2} V_L H_{d\, 1,0} H_{d\, 1,0} m_d V_L^\dagger \right)=0   \\
&\bar{v} B H_{d\, 1,0}-i m_d H_{d\, 2,0}-i H_{d\, 2,0} m_d+\frac{1}{2} m_d H_{d\, 1,0} H_{d\, 1,0} +   \nonumber   \\
& H_{d\, 1,0} m_d H_{d\, 1,0}+\frac{1}{2} H_{d\, 1,0} H_{d\, 1,0} m_d + t_\beta \left( -V_L^\dagger m_u H_{u\, 1,0} V_L.H_{d\, 1,0}\right. -   \nonumber   \\
& V_L^\dagger H_{u\, 1,0} m_u V_L H_{d\, 1,0}+i V_L^\dagger m_u H_{u\, 2,0} V_L+i V_L^\dagger H_{u\, 2,0} m_u V_L   +  \nonumber   \\
& \left. \frac{1}{2} V_L^\dagger m_u H_{u\, 1,0} H_{u\, 1,0} V_L-\frac{1}{2} V_L^\dagger H_{u\, 1,0} H_{u\, 1,0} m_u V_L \right)=0  \\
& -H_{d\, 1,0} H_{R\, 1,0} V_L^\dagger-i H_{d\, 2,0} V_L^\dagger+\frac{1}{2} H_{d\, 1,0} H_{d\, 1,0} V_L^\dagger+i V_L^\dagger H_{u\, 2,0} - \nonumber   \\
& \frac{1}{2} V_L^\dagger H_{u\, 1,0} H_{u\, 1,0}+i H_{R\, 2,0} V_L^\dagger+\frac{1}{2} H_{R\, 1,0} H_{R\, 1,0} V_L^\dagger=0
\end{align}
Finally, taking the coefficient of $k^0\times y^2$ in the Eqs.~\eqref{system}and~\eqref{e3},  one gets three linear equations to be solved for $H_{u\,0,2},H_{d\,0,2},H_{R\,0,2}$:
\begin{align}
& V_L m_d H_{d\, 0,1} V_L^\dagger-V_L H_{d\, 0,1} m_d V_L^\dagger-2 V_L m_d V_L^\dagger H_{u\, 0,1} -   \nonumber   \\
& i m_u H_{u\, 0,2}-i H_{u\, 0,2} m_u+\frac{1}{2} m_u H_{u\, 0,1} H_{u\, 0,1}+H_{u\, 0,1} m_u.H_{u\, 0,1}+\frac{1}{2} H_{u\, 0,1} H_{u\, 0,1} m_u + \nonumber   \\
& t_\beta \left( -V_L m_d H_{d\, 0,1} V_L^\dagger H_{u\, 0,1}-V_L H_{d\, 0,1} m_d V_L^\dagger H_{u\, 0,1}\right. + \nonumber   \\
&  i V_L m_d H_{d\, 0,2} V_L^\dagger+i V_L H_{d\, 0,2} m_d V_L^\dagger+\frac{1}{2} V_L m_d H_{d\, 0,1} H_{d\, 0,1} V_L^\dagger- \nonumber  \\
& \left. \frac{1}{2} V_L H_{d\, 0,1} H_{d\, 0,1} m_d V_L^\dagger\right)=0  \\
& 2 V_L^\dagger m_u V_L H_{d\, 0,1}-V_L^\dagger m_u H_{u\, 0,1} V_L+V_L^\dagger H_{u\, 0,1} m_u V_L-i m_d H_{d\, 0,2} -   \nonumber  \\
&  i H_{d\, 0,2} m_d+\frac{1}{2} m_d H_{d\, 0,1} H_{d\, 0,1}+H_{d\, 0,1} m_d H_{d\, 0,1}+\frac{1}{2} H_{d\, 0,1} H_{d\, 0,1} m_d +  \nonumber  \\
& t_\beta \left( V_L^\dagger m_u H_{u\, 0,1} V_L H_{d\, 0,1}-V_L^\dagger H_{u\, 0,1} m_u V_L H_{d\, 0,1}+i V_L^\dagger m_u H_{u\, 0,2} V_L \right. +   \nonumber  \\
& \left. i V_L^\dagger H_{u\, 0,2} m_u V_L+\frac{1}{2} V_L^\dagger m_u H_{u\, 0,1} H_{u\, 0,1} V_L-\frac{1}{2} V_L^\dagger H_{u\, 0,1} H_{u\, 0,1} m_u V_L \right) =0
\end{align}
\begin{align}
& H_{d\, 0,1} H_{R\, 0,1} V_L^\dagger+i H_{d\, 0,2} V_L^\dagger-\frac{1}{2} H_{d\, 0,1} H_{d\, 0,1} V_L^\dagger-i V_L^\dagger H_{u\, 0,2} + \nonumber  \\
& \frac{1}{2} V_L^\dagger H_{u\, 0,1} H_{u\, 0,1}-i H_{R\, 0,2} V_L^\dagger-\frac{1}{2} H_{R\, 0,1} H_{R\, 0,1} V_L^\dagger=0   \,.
\end{align}
Once all these $H_{u,d},H_R$ are known, one reconstructs the matrices $S_{u,d}$ and $V_R$ up to the second order in both $k$ and $y$, by using the definitions in Eqs.~\eqref{definition} and~\eqref{e3}.

\bibliographystyle{jhep}
\bibliography{biblio}

\end{document}